\documentclass[authoryear,preprint,review,12pt]{elsarticle}

\usepackage{lineno,hyperref}
\modulolinenumbers[5]

\usepackage{color}
\usepackage{graphicx}

\usepackage{latexsym}
\usepackage{amsmath}
\usepackage{amssymb}
\usepackage{amsfonts}
\usepackage{bm}

\usepackage{graphicx}
\usepackage{graphics}
\usepackage{caption}
\usepackage{subcaption}
\usepackage{float}
\usepackage{placeins} 
\usepackage{lastpage}
\usepackage[format=plain,justification=raggedright,singlelinecheck=false,font=small,labelfont=bf,labelsep=space]{caption}

\usepackage{array}
\usepackage{xmpmulti}
\usepackage{multirow}

\journal{Solar energy}

\biboptions{authoryear}

\begin{document}

\begin{frontmatter}



\title{An empiric-stochastic approach, based on normalization parameters, to simulate solar irradiance}



\author[label1]{Edith Osorio de la Rosa}
\author[label1]{Guillermo Becerra Nu\~nez}
\address[label1]{CONACYT, Universidad de Quintana Roo, Blv. Bahía s/n esq. Ignacio Comonfort, C.P. 77019, Quintana Roo,México.}
\author[label2]{Alfredo Omar Palafox Roca}
\author[label2,label3]{Ren\'{e} Ledesma-Alonso}
\address[label2]{Universidad de las Am\'{e}ricas Puebla, Sta. Catarina, San Andr\'{e}s Cholula, C.P. 72810, Puebla, M\'{e}xico.}
\address[label3]{e-mail: rledesmaalonso@gmail.com}

\begin{abstract}
The data acquisition of solar radiation in a locality is essential for the development of efficient designs of systems, whose operation is based on solar energy.
This paper presents a methodology to estimate solar irradiance using an empiric-stochastic approach, which consists on the computation of normalization parameters from solar irradiance data.

For this study, solar irradiance data was collected with a weather station during a year.
Post-treatment included a trimmed moving average, to smooth the data, the performance a fitting procedure using a simple model, to recover normalization parameters, and the estimation of a probability density map by means of a kernel desnsity estimation method.
The normalization parameters and the probability density map allowed us to build an empiric- stochastic methodology that generates an estimate of the solar irradiance.
In order to validate our method, simulated solar irradiance has been used to compute the theoretical generation of solar power, which in turn has been compared to experimental data, retreived from a commercial photovoltaic system.
Since the simulation results show a good agreement has been with the experimental data, this simple methodology can estimate the solar power production and may help consumers to design and test a photovoltaic system before installation.
\end{abstract}

\end{frontmatter}

\section{Introduction}

Solar radiation data provide information on how much of the energy provided by the sun is irradiated on the earth's surface, during a particular time period. 
This data is essential for the design, cost and energy analysis of solar power systems, and the optimal exploitation of solar energy [\cite{Wittmann2008,MARTIN20101772}].
In particular, an improvement of the solar irradiance forecast would be very useful in predicting the performance of photovoltaic systems [\cite{LI2016542,Ibrahim2017,Bora2018}].
Nevertheless, the measurement of global horizontal radiation (GHI) data is only available in specific regions, since the initial and maintainance cost of weather stations are high.
Therefore, solar energy simulation techniques stand out among the available solutions to generate GHI data on any earth's place, due to its relative low cost [\cite{Weiss2004,Wan2015}].
This important advantage is extended by a significant attribute of simulations, \emph{i.e.} they can provide data at any discrete time step, when the appropriate model is employed.
For instance, data acquired during an hour can make a significant difference in the design, evaluation and improvement of solar energy systems [\cite{YANG20123531,Jahani2017,Aoun2017}].

Another situation arises when solar irradiance data is obtained with short time-steps.
Meteorological or ambient conditions, such as sky blockages, relative humidity and ambient temperature, hinder the correct measurement of the full and clear sky solar irradiance [\cite{GUEYMARD20122145}]. 
However, these adverse conditions must be taken into account to provide accurate and more realistic predictions of the electrical charge or the thermal energy generated by solar power systems [\cite{DIAGNE201365}]. 

On the other hand, several techniques had been developed for the determination of the amount of solar irradiance incident on the Earth's surface [\cite{Khatib2012,Wan2012}], with the purpose of improving the accuracy of the predictions.
Among them, time series analysis [\cite{MARTIN20101772,YANG20123531}] and artificial neural networks [\cite{Inman2013,Kaushika2014,Gutierrez2016}] are the preferred approaches, when enough historical data is available.
As well, empirical models that estimate solar radiation have been developed, offering low computational costs and great accuracy [\cite{Belcher2007,Nwokoloi2017Renew}].
Additionally, the use of hybrid models has gained popularity, since they increase the accuracy of the predictions, by taking advantage of the strengths of the differents models [\cite{Inman2013,Gordon2009,Ji2011}].

This article is aimed to explain the implementation of an empiric-stochastic methodology to generate solar irradiance data.
From experimental measurements, the method is based on the computation of normalization prarameters, average trends and probability distributions, which are used to perform simulations of the solar irradiance along a day.
In Section~\ref{Sec:Method}, a detailed description of the methodology is provided.
In Section~\ref{Sec:Results}, the results from the application of the methodology, departing from experimental measurements of solar irradiance, are presented and analysed.
A description of the normalization parameters, average trends and probability distributions, computed for the measured data, is given. 
In Section~\ref{Sec:Sims}, performed simulations of solar irradiance are compared with experimental data, in two different ways: 1) a direct comparison concerning solar irradiance and irradiant exposure; 2) an indirect comparison refering to solar-cell electric charge which requires the conversion of the simulated irradiance into electric charge, by means of a photovoltaic model. 
In Section~\ref{Sec:Conclusions}, some final commments about the developed methodology and its employment are discussed.

\section{Methodology: Data acquisition and post-treatment}
\label{Sec:Method}

The four calendar-based seasons are traditionally bounded by the solstices and equinoxes.
For instance, the 2017-2018 seasons were: Spring from March 21st, 2017 (spring equinox)  to June 21st, Summer from June 22st (summer solstice) to September 21st, Autumn from September 22st (autumn equinox) to December 20st, and Winter from December 21st (winter solstice) to March 20st, 2018.
Herein we have decided to consider four non-traditional seasons, which are presented and described in Table~\ref{Tab1}.
This convention has been chosen, since there is more likeness between the dates around a solstice or an equinox, from an astronomical view-point, than there is for the dates going from a solstice to an equinox, or vice versa.

\begin{table}[h]
\renewcommand\arraystretch{1.5}
\centering
\caption{Astronomical seasons definition employed in this work.}
\begin{tabular}{c|cc|cc}
Season & \multicolumn{2}{c|}{from day} & \multicolumn{2}{c}{to day} \\ \hline
1 & 35 & (February 4, 2017) & 124 & (May 4, 2017) \\ \hline
2 & 125 & (May 5, 2017) & 218 & (August 6, 2017) \\ \hline
3 & 219 & (August 7, 2017) & 309 & (November 5, 2017) \\ \hline
4 & 310 & (November 6, 2017) & 34 & (February 3, 2018) \\ \hline
\end{tabular}
\label{Tab1}
\end{table}

\begin{figure}[!h]
\centering
\includegraphics[width=\textwidth]{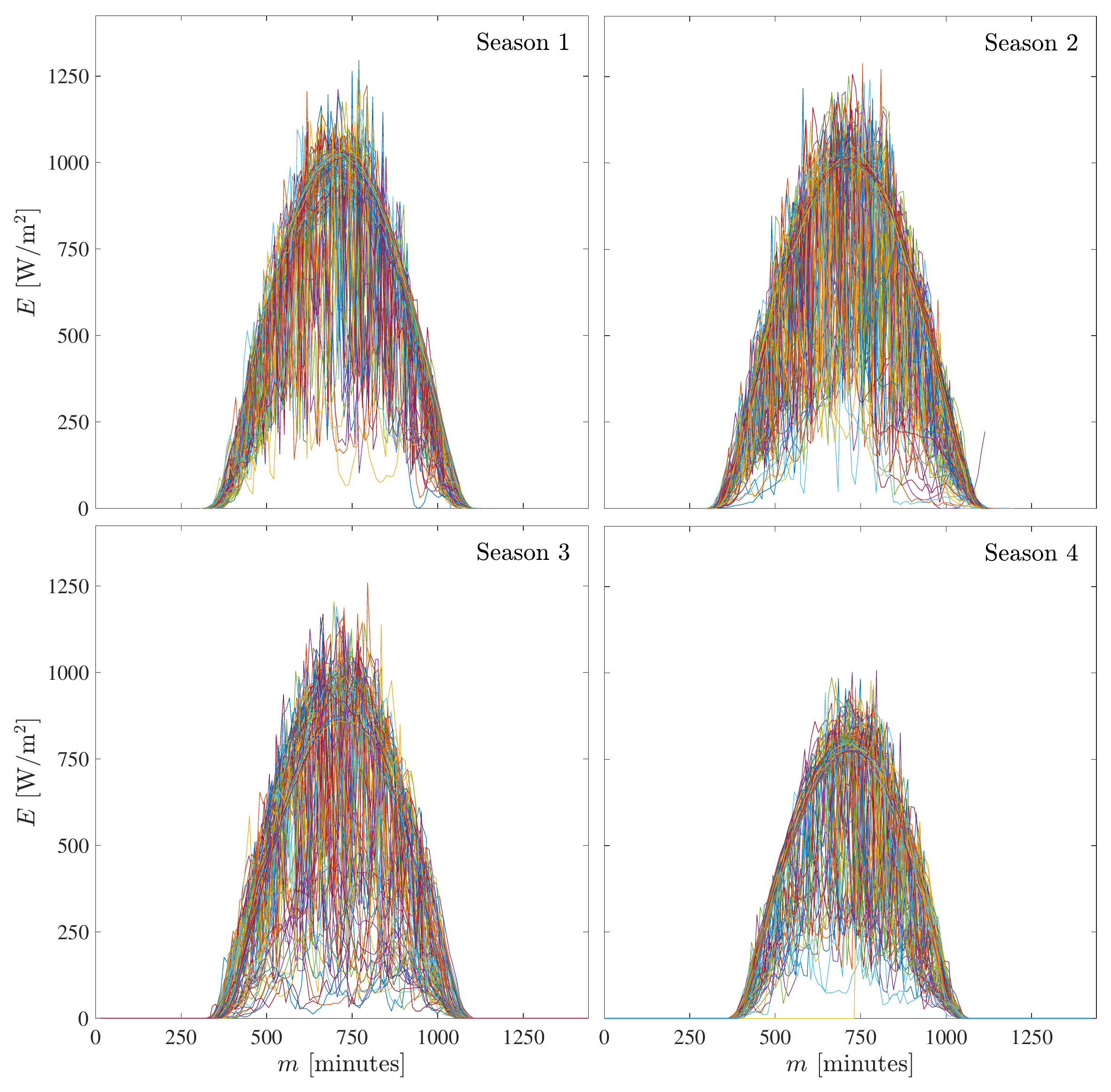} 
\caption{Experimental data of solar irradiance (GHI), denoted by the variable $E$, as a function of the short-term variable $m$, for the 2017-2018 seasons described in Table~\ref{Tab1}.}
\label{Fig1}
\end{figure}

The campus Chetumal of the Universidad de Quintana Roo (UQRoo-Chetumal) is located at the city of Chetumal, on the southern coast of the Yucatan Peninsula, in the mexican state of Quintana Roo.
The weather station of the UQRoo-Chetumal is placed on the roof of building F, ``Ing. Luis Felipe Leyva'', Engineering Workshop.
Its geographic coordinates are 18$^{\circ}$31'25'' N and 88$^{\circ}$16'7'', in latitude and longitude, respectively.
\\
The GHI was measured with a thermopile pyranometer model 8-48, collecting data from February to August 2017 under all sky conditions.
Hereafter, GHI will be called simply the irradiance $E(t)$, which is a function of time $t$, and the reference time $t_0=0$ is taken to be midnight on (March 21st) January the 1st, 2017.
For simplicity, the temporal variable is decomposed as follows:
\begin{equation}
t=1440d+m \ ,
\end{equation}
where $t$ is the time in minutes with respect to $t_0$, $d$ indicates the day of the year (long-term variable) and $m$ denotes the elapsed time in minutes with respect to midnight on each day (short-term variable).
Therefore, the irradiance can be now defined in terms of two time variables, i.e. $E(t)\rightarrow E(d,m)$, which accounts for the two different time scales previously described.
Due to this decomposition, the irradiance data of all the days of each season can be plotted and overlapped in a single graphic, as it is shown in Figure~\ref{Fig1} for the four seasons.
Additionally, notice that the irradiance data has been obtained along the daytime, with a sampling frequency of 1$/$10 $min^{-1}$.
The irradiance curves, together with the results of our methodology and analysis, are discussed in Section~\ref{Sec:Results}.

\subsection{Trimmed moving average}

\begin{figure}[!h]
\centering
\includegraphics[width=\textwidth]{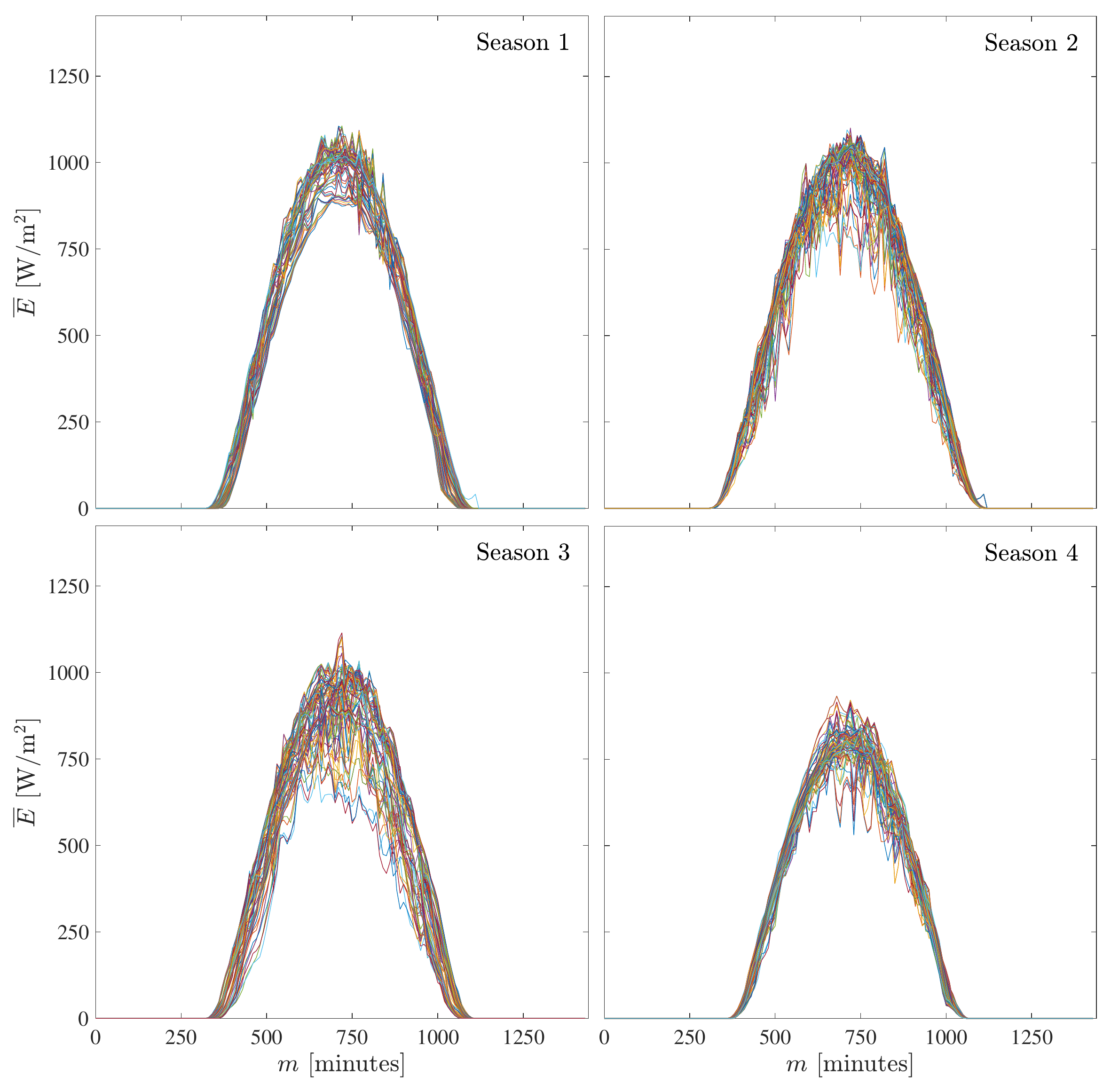} 
\caption{Irradiance data shown in Figure~\ref{Fig1} after the application of the TMA, as described by eq.~\eqref{Eq:TMA} with $N=5$ and $L=4$, denoted by the variable $\overline{E}$, as a function of the short-term variable $m$, for the 2017-2018 seasons described in Table~\ref{Tab1}.}
\label{Fig2}
\end{figure}

In order to estimate the clear sky irradiance (CSI), a trimmed moving average (TMA) has been proposed.
This approach consists in averaging the symmetric data, from $N$ days before to $N$ days after the date of interest $d$, excluding the $L$ lowest values or outliers.
This TMA is given by:
\begin{align}
\overline{E}\left(d,m\right)=\left(\dfrac{1}{2N+1-L}\right)\left\{\sum_{n=-N}^N E\left(d+n,m\right)-\sum_{l\in \mathcal{L}} E\left(l,m\right) \right\} \ ,
\label{Eq:TMA}
\end{align}
where $2N+1-L$ is the number of days employed to perform this averaging process, and the day $l\in\left[d-n,d+n\right]$ belongs to the set $\mathcal{L}$, formed by the $L$ days that present the lowest irradiance.
The application of this numerical procedure smooths out short-term fluctuations, originating from strange meteorological conditions or instrument errors-deletions, and highlights long-term irradiance trends, the diffuse and global irradiance at the local position.
The TMA has been applied to the data presented in Figure~\ref{Fig1}, with $N=5$ and $L=4$, yielding the smoothed curves displayed in Figure~\ref{Fig2}.
It is important to emphasize that the objective of the TMA step in our methodology is not to replace the data, but to obtain more regular curves on which the application of fitting, normalizing and probability estimation procedures should be straightforward.

\subsection{Normalization variables and master curve}

\begin{figure}[!h]
\centering
\includegraphics[width=\textwidth]{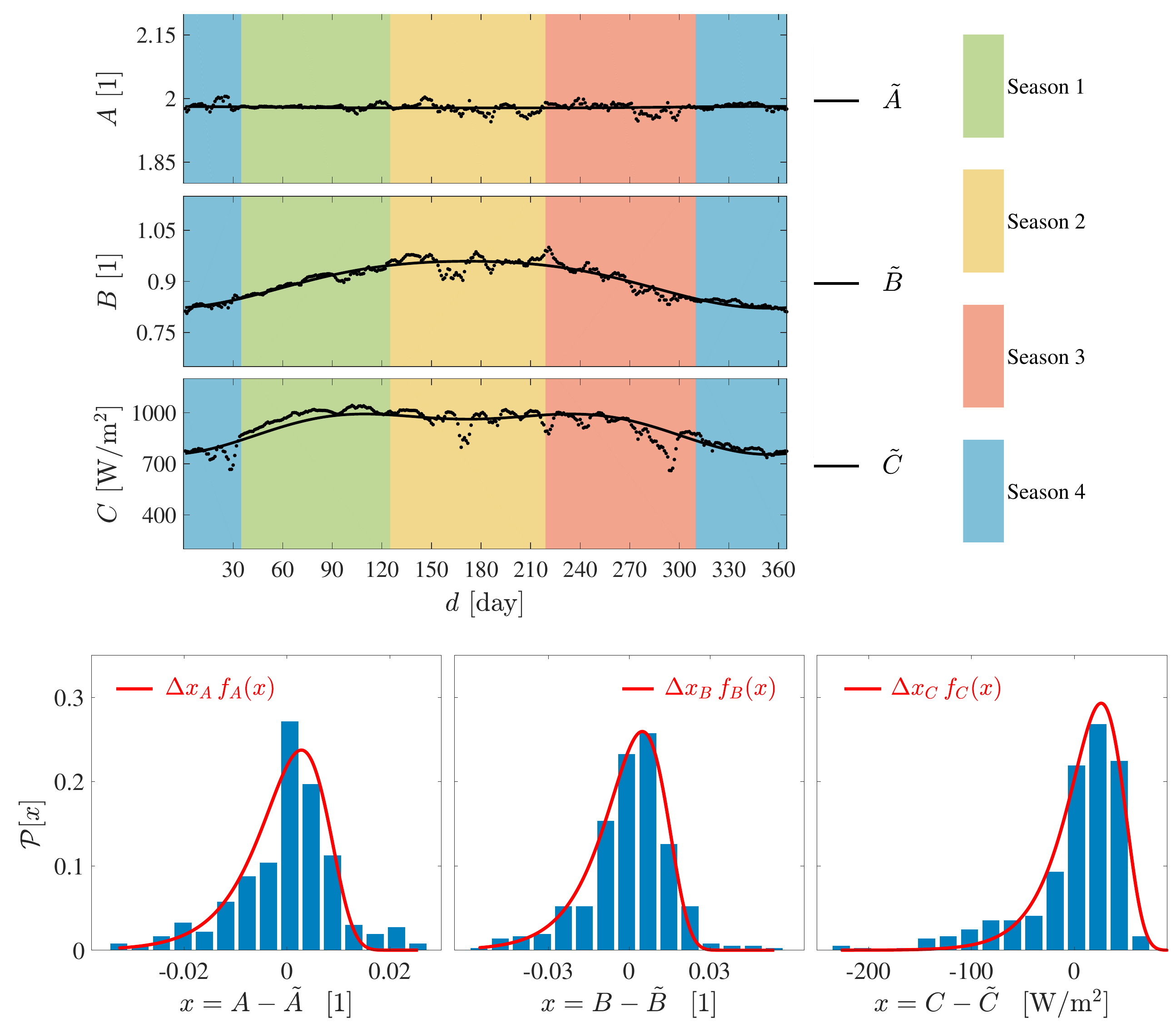} 
\caption{\textbf{Top}: Evolution of the three fitting parameters $A$, $B$ and $C$ (dots) and their general trends $\tilde{A}$, $\tilde{B}$ and $\tilde{C}$ (solid lines), in terms of the long-term variable $d$.
The parameters were introduced in eq.~\eqref{Eq:Fit}, the general trends in eqs.~\eqref{Eq:Trends} and their physical interpretation is given in text.
\textbf{Bottom}: Histograms of the residuals $x$, the difference between each fitting parameter and its corresponding trend, and the proposed probability distribution function (solid lines), which are scaled by the bins width for comparison purposes. 
}
\label{Fig3}
\end{figure}

As it is shown in Figure~\ref{Fig1}, the daily behavior of the irradiance $E$, as a function of the short-term variable $m$, is mostly represented by a bell-like shape.
Therefore, the natural choice has been to fit the data with a parabola, in order to decouple the dependence on the long-term variable $d$ and the short-term variable $m$, and to obtain a simple expression to describe the irradiance $\overline{E}$ as a function of both variables.
The parabolic fitting curve reads:
\begin{equation}
\overline{E}\left(d,m\right)=C\left(d\right)\left\{1-\left[\dfrac{1}{B\left(d\right)}\right]^2\left[\dfrac{m}{m_c}-A\left(d\right)\right]^2\right\} \ ,
\label{Eq:Fit}
\end{equation}
where $m_c$ is a characteristic value of the short time-scale, which we define as one half-daytime at the equator $m_c=360$ minutes, and the parameters $A$, $B$ and $C$ depend on the long-term variable $d$.
At the given latitude, $A\left(d\right)$ indicates the time, in terms of multiples of $m_c$, at which noon occurs, $B\left(d\right)$ denotes the half-daytime, also in terms of multiples of $m_c$, and $C\left(d\right)$ is the irradiance at noon, the power per unit area in W$/$m$^2$.
The evolution of the three fitting parameters, as functions of the long-term variable $d$, are shown in Figure~\ref{Fig3}.
Since these parameters are closely related to the periodic motion of the earth around the sun, we expect them to show periodic behaviors with frequency $2\pi/365$ and to be in phase with the summer solstice.
These ideas, together with a detailed description of the parameters trend, are discussed in Section~\ref{Sec:Results}.

\begin{figure}[!h]
\centering
\includegraphics[width=0.9\textwidth]{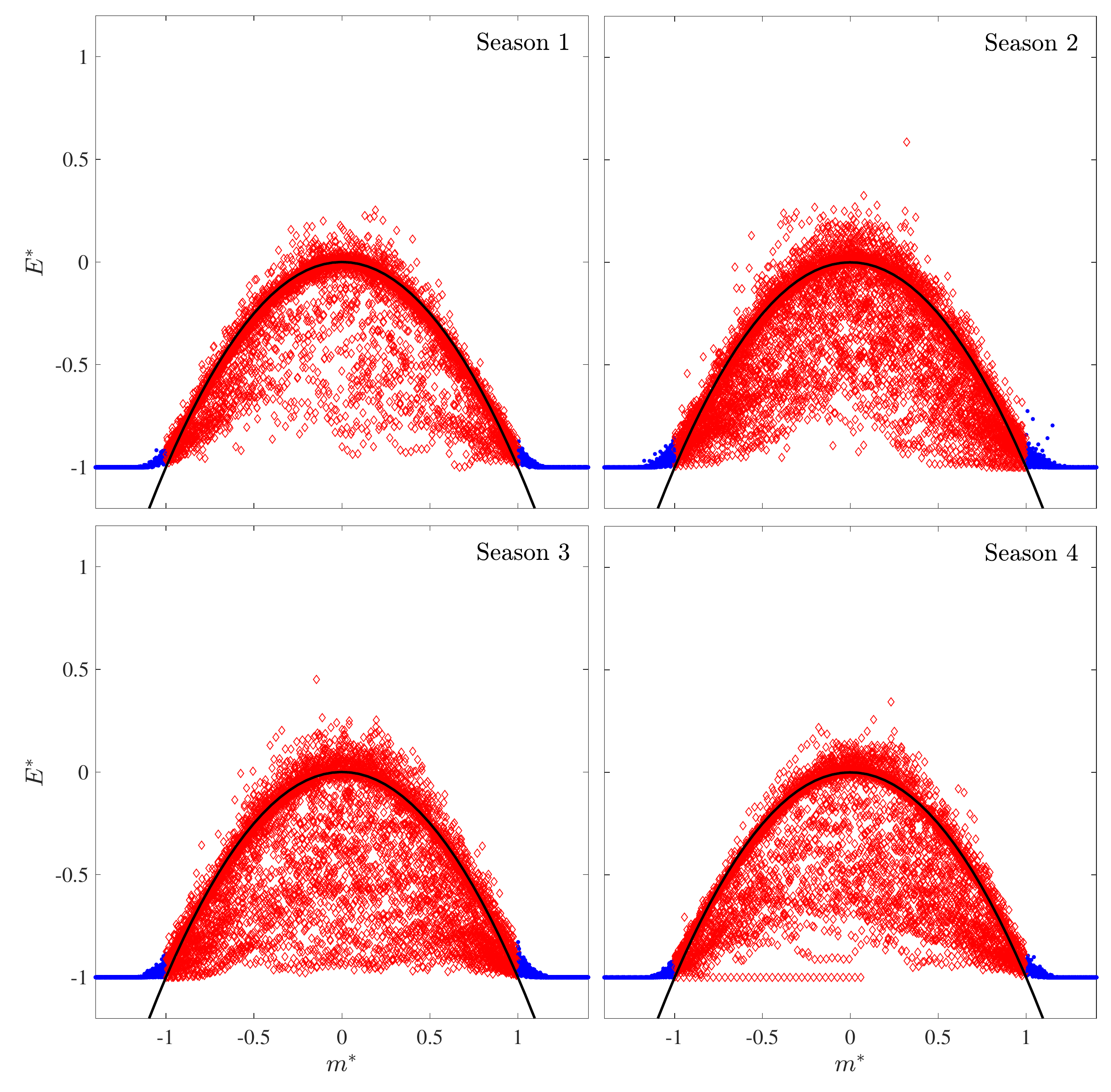} 
\caption{
Normalized irradiance $E^{\ast}$ as a function of the normalized short-term variable $m^{\ast}$, defined in eq.~\eqref{Eq:NormVar}.
The results from the experimental data is shown as diamonds (red) if $m^{\ast}$ lies within the daytime limits $m_{0\pm}$, whereas dots (blue) are employed for the data outside the aforesaid boundaries.
The irradiance estimator $\mathcal{E}$ (black line), given in eq.~\eqref{Eq:Master}, is presented as well.}
\label{Fig4}
\end{figure}

The times of sunrise and nightfall at the given latitude can be approximated with the following expression, obtained as the time $m=m_{0\pm}$ of zero irradiance $\overline{E}=0$ from eq.~\eqref{Eq:Fit}:
\begin{equation}
m_{0\pm}=m_c\left(A\pm B\right) \ .
\label{Eq:Lims}
\end{equation}
The negative subindex indicates sunrise, whereas the positive subindex indicaes nightfall.
The analysis in this manuscript will be focused on the irradiance data occuring within the range $m\in\left[m_{0-},m_{0+}\right]$.
\\
Now, using the information obtained from the fitting parameters, one may procede to normalize the original data (before the application of the TMA), with the purpose of finding a general trend for the global irradiance.
The following normalization variables:
\begin{align}
E^{\ast}\left(m^{\ast}\right)&=\dfrac{E\left(d,m\right)}{C\left(d\right)}-1 \ , & m^{\ast}&=\dfrac{1}{B\left(d\right)}\left[\dfrac{m}{m_c}-A\left(d\right)\right] \ ,
\label{Eq:NormVar}
\end{align}
make the irradiance/time data collapse into a general trend, which, in turn, may be described by the irradiance estimator $\mathcal{E}$ defined by the expression:
\begin{equation}
\mathcal{E}\left(m^{\ast}\right)=-\left(m^{\ast}\right)^2 \ ,
\label{Eq:Master}
\end{equation}
which forecasts a general behavior of the normalized irradiance.
As a consequence, the normalized daytime is $m^{\ast}\in\left[m_{0-}^{\ast},m_{0+}^{\ast}\right]$, where limits are $m_{0\pm}^{\ast}=\pm 1$, respectively. 
The normalization, described by eq.~\eqref{Eq:NormVar}, has been applied to the irradiance data and a comparison with the irradiance estimator $\mathcal{E}$, given in eq.~\eqref{Eq:Master}, is presented in Figure~\ref{Fig4} as function of the normalized short-term variable $m^{\ast}$, for each season defined in Table~\ref{Tab1}.
The estimator $\mathcal{E}$ seems to fairly describe the average trend of the normalized experimental data $E^{\ast}$, since the data density is higher around the parabolic shape of $\mathcal{E}$.
More details about the normalization effects can be found in Section~\ref{Sec:Results}.

\subsection{Residuals, discrete probability map and probability distribution map}
\label{subsec:Res}

\begin{figure}[!h]
\centering
\includegraphics[width=\textwidth]{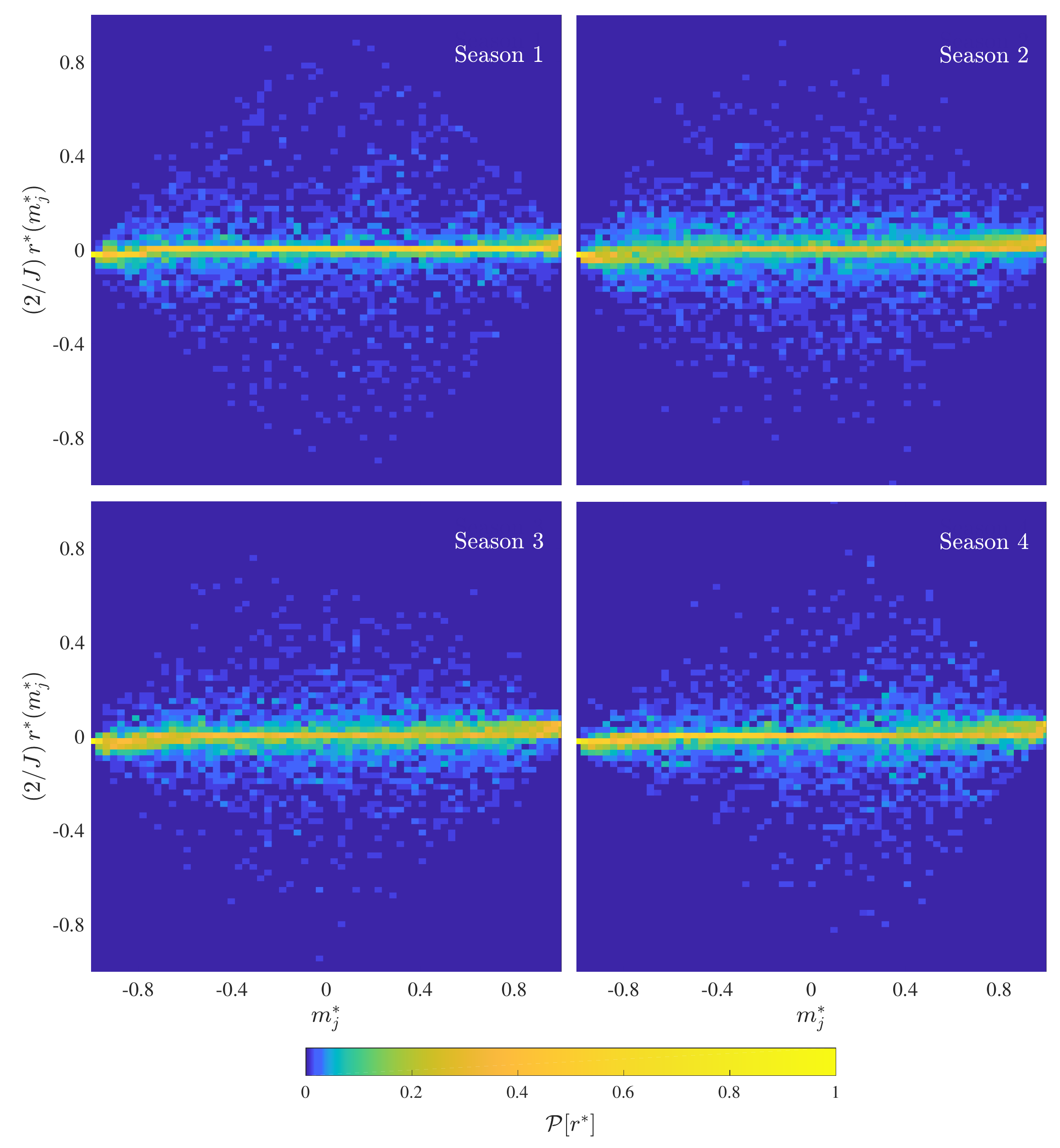} 
\caption{Discrete probability maps, in terms of the discrete short-term variable $m^{\ast}_j$ (horizontal axis) and the corresponding to the rate of change of the residuals $\displaystyle r^{\ast}\left(m^{\ast}_j\right)$.}
\label{Fig5}
\end{figure}

Despite the affinity between the estimator $\mathcal{E}$ and the average trend of the normalized experimental data $E^{\ast}$, a significant dispersion cannot be neglected.
To quantify this dispersion, the residuals between the normalized experimental data $E^{\ast}$ and the irradiance estimator $\mathcal{E}$ are computed as:
\begin{equation}
R^{\ast}\left(m^{\ast}\right)=E^{\ast}\left(m^{\ast}\right)-\mathcal{E}\left(m^{\ast}\right) \ .
\label{Eq:ERes}
\end{equation}
where $R^{\ast}$ may be considered as a random variable, which clearly separates $E^{\ast}$ from its estimator $\mathcal{E}$.
Nevertheless, in the course of the present analysis, we have realized that the evolution of the residuals is better described by a sequential variation.
This is due to the fact that the evolution of the solar irradiance is correlated from one instant to another, which is noticeable in our data, despite the 10 minutes lapse between consecutive acquisition moments.
Therefore, the residuals $R^{\ast}\left(m^{\ast}\right)$ may not be taken as random independent variables.
We should consider a \emph{second order model}, for which the rate of change of the residuals $r^{\ast}=d\, R^{\ast}/d\, m^{\ast}$ becomes the independent random variable.
Following this statement, we define the normalized residual at a given value of the short-term variable with the expression:
\begin{align}
R^{\ast}\left(m^{\ast}\right)&=R^{\ast}\left(-1\right)+\int_{-1}^{m^{\ast}} r^{\ast} d m^{\ast} &
\text{with} \quad r^{\ast}=\dfrac{d\, R^{\ast}}{d\, m^{\ast}} \ ,
\label{Eq:resdif}
\end{align}
As it can be observed in Figure~\ref{Fig4}, the computation of the residuals only makes sense within the normalized daytime region $m^{\ast}\in\left[-1,1\right]$.
Additionally, one can notice that it is difficult to identify the exact number of points concentrated by zones, which consequently may provoke the same effect on the residuals $R^{\ast}$ and its rate of change $r^{\ast}$.
Therefore, a discretization $m^{\ast}_j$ of the dimensionless short-term variable $m^{\ast}$ is proposed, in order to enumerate the frequency of the residuals rate of change and yield a histogram for each discrete value of $m^{\ast}_j$ with $j=0,1,2,\dots J$.
The values of $m^{\ast}_j$ are spaced by $2/J=0.0313$ dimensionless units, which corresponds approximately to 10.07 minutes, whereas its first and last values should be $-1$ and $1$, respectively. 
In other words, we propose the discretization given by:
\begin{equation}
m^{\ast}_j=-1+\dfrac{2\, j}{J} \quad \text{with } j=0,1,2,\dots J \quad \text{and } J=\left\lfloor\dfrac{m_c\langle B\rangle}{5}\right\rfloor\ ,
\label{Eq:Discrete}
\end{equation}
where $\langle B \rangle=0.8990$ is the average value of the half-daytime $B$ and the number of discrete values of $m^{\ast}_j$ is $J+1=65$.
All the histogram are computed using the same number of bins $M_r$, which is the average of the required number of bins for each histogram $M_{r,j}$, computed with the Freedman Diaconis' rule [\cite{Freedman1981}] and with positions in the range $r^{\ast}\in[-J/2,J/2]$, which encloses properly the values of the rate of change of the residuals.
Once $M_r$ is known, one may proceed to determinate the bin width $\Delta r^{\ast}$.
In brief, the above-mentioned methodology is summarized by the following formulas:
\begin{subequations}
\begin{align}
\Delta r^{\ast}\left(m^{\ast}_j\right)&=2\, \dfrac{\text{IQR}\left[r^{\ast}\left(m^{\ast}_j\right)\right]}{\left[\ell\left(m^{\ast}_j\right)\right]^{1/3}} &
M_{r,j}&=\left\lceil\dfrac{\max\left[r^{\ast}\left(m^{\ast}_j\right)\right]-\min\left[r^{\ast}\left(m^{\ast}_j\right)\right]}{\Delta r^{\ast}\left(m^{\ast}_j\right)}\right\rceil \ , \label{Eq:FDrule}\\
M_r&=\dfrac{1}{J}\sum^{J}_{j=1}M_{r,j} &
\Delta r^{\ast}&=\left\lceil\dfrac{max(r^{\ast})-min(r^{\ast})}{M_r}\right\rceil \ , \label{Eq:BinAv}
\end{align}
\end{subequations}
where $\ell\left(m^{\ast}_j\right)$ is the number of $r^{\ast}\left(m^{\ast}_j\right)$ samples and $\text{IQR}\left[\cdot\right]$ indicates the interquartile range of the data, at a given value of the short-term variable $m^{\ast}_j$ and in the trimmed range $r^{\ast}\in[-J/2,J/2]$.
The equations in \eqref{Eq:FDrule} describe the Freedman Diaconis' rule, whereas equations in \eqref{Eq:BinAv} indicate simple averaging and bin determination formulas.
For this case, we have found the value of $\Delta r^{\ast}=0.0244$ and $M_r=85$.
Finally, each histogram must be normalized to build up a discrete probability map, whose intensity (normalized frequency) describes the probability of a rate of change for the residuals $r^{\ast}$ at each value of the dimensionless short-term variable $m^{\ast}$.
This procedure for obtaining a discrete probability map for the rate of change of the residuals $r^{\ast}$ was applied on each season, as depicted in Figure~\ref{Fig5}.

\begin{figure}[!h]
\centering
\includegraphics[width=\textwidth]{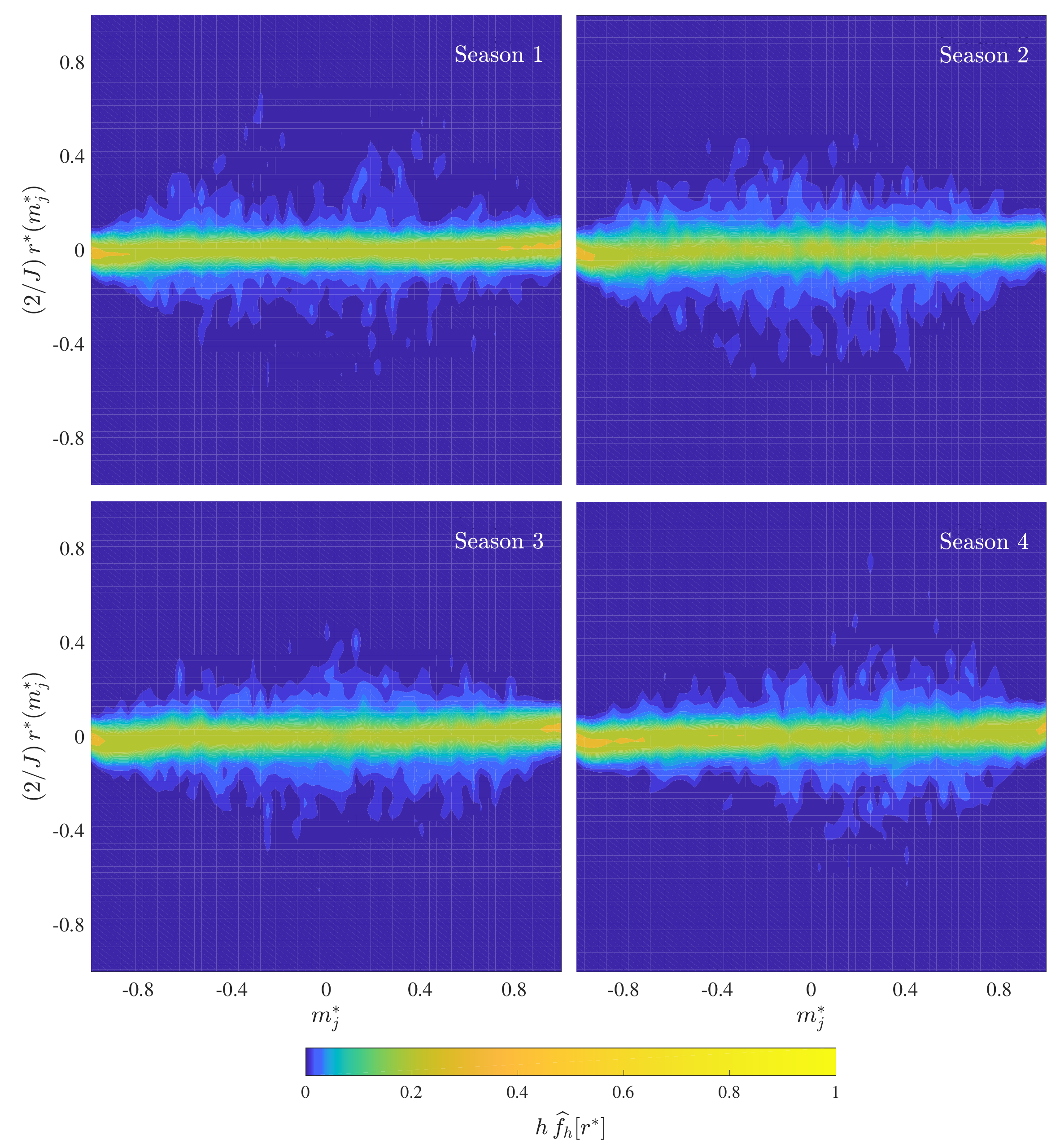} 
\caption{Scaled probability density maps (Scaled-PDMs), in terms of the discrete short-term variable $m^{\ast}_j$ (horizontal axis) and the corresponding to the rate of change of the residuals  $\displaystyle r^{\ast}\left(m^{\ast}_j\right)$. A bandwith value of $h=0.0364$ has been employed, which justification is provided in the text.}
\label{Fig6}
\end{figure}

Additionally, in order to estimate a probability density map (PDM) from the data, a kernel density estimation (KDE) method [\cite{Silverman}] was employed.
This method, applied to the rate of change of the residuals $r^{\ast}$, consists in the calculation of the following density estimator:
\begin{align}
\widehat{f}_h\left(r^{\ast}\right)=\dfrac{1}{h\, S}\sum_{s=1}^S K\left(\dfrac{r^{\ast}-r^{\ast}_s}{h}\right) && \text{for } -\infty<r^{\ast}<\infty\ ,
\end{align}
where $r^{\ast}_s$ are the discrete samples, with $1\leq s\leq S$ and $S$ being the total number of samples, $h$ is the bandwidth and $K(\cdot)$ is the kernel, which has been chosen to be a Gaussian function with mean $\mu=0$ and variance $\sigma^2=1$ reading:
\begin{equation}
K\left(u\right)=\dfrac{1}{\sqrt{2\pi}}\exp\left(\dfrac{-u^2}{2}\right)\ .
\label{Eq:Kernel}
\end{equation}
The bandwidth $h$ has been set to the maximum value of $h_j$, according to the discretization of $m^{\ast}_j$, with $j=0,1,2,\dots,J$.
Using a normal distribution approximation, the following formulas are applied:
\begin{align}
h&=\langle h_j\rangle &  h_j=\left(\frac{4}{3\, \ell\left(m^{\ast}_j\right)}\right)^{\frac{1}{5}}\hat{\sigma}\left[r^{\ast}\left(m^{\ast}_j\right)\right] \ ,
\end{align}
where $\langle h_j \rangle$ is the average value of $h_j$ and $\hat{\sigma}\left(m^{\ast}_j\right)$ is the standard deviation of the $r^{\ast}\left(m^{\ast}_j\right)$ samples.
This procedure for obtaining a PDM was applied on each of the four seasons described in Table~\ref{Tab1}, as shown in Figure~\ref{Fig6}.
The value of $h$ was obtained taking into account the residuals of the 365 days of measurements, thus yielding a single value $h=0.0364$ employed to compute, homogeneously, the PDMs for all the seasons.

A complete discussion on the rate of change of the residuals and a detailed description of the discrete probability and probability density maps is given in Section~\ref{Sec:Results}.

\section{Results}
\label{Sec:Results}

\subsection{Short-term solar irradiance}

The GHI data acquired for each season is presented in Figure \ref{Fig1}.
In the four cases, the global irradiance data shows the superposition of a bell shape and a noise signal.
The bell shape suggestes an average behavior, whereas the noise is the consequence of any sky blockage, \emph{e.g.} clouds and fleeting shadows, or meteorological condition, \emph{e.g.} humidity, that decrease the GHI, and temporary reflecting surfaces or clear sky conditions that increase the GHI.
The sunrise, between 5:30 AM (300 minutes) and 6:30 AM (390 minutes), and the sunset, between 5:30 PM (1050 minutes) and 6:30 AM (1110 minutes), are clearly observed and vary according to the season, as a consequence of earth's orbital movement [\cite{Vieira2012}].
The sunrise and sunset are indicated by the change from or towards, respectively, a cero value of the solar irradiance.
The maxima of the GHI, which show values around 1000 W$/$m$^2$ for Seasons 1-3 and lower values  around 800 for Season 4, occur in the range of the short-term variable $m\in[600,800]$ minutes.
Some minutes apart from the sunrise and sunset instants, a parabolic shape is well defined for clear-sky days, where half-daytime occurs at $m=700$ minutes.
For cloudy days, the irradiance lessens to values below 250 W$/$m$^2$.
A TMA was applied to the acquired data, resulting in smoothed curves $\overline{E}(m)$ which are shown in Figure \ref{Fig2}.
This procedure removes the effect of sky blockage and yields curves that highlight the aforementioned bell shape of the GHI curves.
These smoothed results allow us to compute the normalization parameters for each day, introduced in eq.~\eqref{Eq:Fit} and discussed in the following subsection.
Using these parameters, we describe the GHI average trend by means of a normalized curve $E^{\ast}(m^{\ast})$ for each day, according to its corresponding season, with the definition of the dimensionless variables given by eq.~\eqref{Eq:NormVar}.
In Figure \ref{Fig4}, the normalized irradiance data is compared to the master curve, a parabola described in eq.~ \eqref{Eq:Master}.
Even though most of the data points overlap in the surroundings of the parabolic trend, a significant dispersion is in sight.
A detailed analysis of this dispersion is re-engaged after the following discussion about the normalization parameters and their behavior.

\subsection{Long-term solar irradiance}

As introduced in the previous subsection, the normalization parameters $A(d)$, $B(d)$ and $C(d)$, described by eq.~\eqref{Eq:Fit}, were obtained for each day.
The evolution of the noon time $A(d)$, the half-daytime $B(d)$ and the peak irradiance $C(d)$ is presented in Figure~ \ref{Fig3}, in terms of the long-term variable $d$.
Additionally, the background color in the figure indicates the corresponding season.
As it has been already mentioned, the values irradiance data depends on the meteorological conditions and temporary blocking/reflecting surfaces, which provokes a variation of the normalization parameters for each day.
Nevertheless, due to the periodic motion of the earth within the solar system, the normalization parameters must be periodic functions with a frequency of $2\pi/365$ cycles per day, as discussed before in this article.
It is also natural to consider that the peak irradiance occur at the summer solstice (day 172), and we expect that the long-term parameters $Y(d)=\left\{A(d),B(d),C(d)\right\}$ will show periodic behaviors and will be in phase with the summer solstice.
Thus, the trends $\tilde{Y}(d)=\left\{\tilde{A}(d),\tilde{B}(d),\tilde{C}(d)\right\}$ of the three fitting parameters should be accurately described by Fourier series.
We have noticed that a constant term and two cosine harmonic functions are enough, reading:
\begin{align}
\tilde{Y}(d)&=y_0+y_1\cos\left(\dfrac{2\pi}{365}\left[d-172\right]\right)+y_2\cos\left(\dfrac{4\pi}{365}\left[d-172\right]\right)
\label{Eq:Trends}
\end{align}
where the coefficients $y_l$ with $l=0,1,2$, which were obtained by fitting each parameter $Y(d)$ with the trend function $\tilde{Y}(d)$, are given in Table~\ref{Tab2}.
The trend functions $\tilde{Y}(d)=\left\{\tilde{A}(d),\tilde{B}(d),\tilde{C}(d)\right\}$ are also shown in Figure~\ref{Fig3}, as functions of the long-term variable $d$.

\begin{table}[h!]
\renewcommand\arraystretch{1.5}
\centering
\caption{Coefficients of the long-term variable trends, to be used in Eq.~\ref{Eq:Trends}.}
\begin{tabular}{c|c|c|c}
$\tilde{Y}$ & $y_0$ & $y_1$ & $y_2$ \\ \hline
$\tilde{A}$ & 1.9790 & -0.0017 & 0.0005 \\ \hline
$\tilde{B}$ & 0.8990 & 0.0689 & -0.0089 \\ \hline
$\tilde{C}$ & 913.0363 & 103.6416 & -54.6980 \\ \hline
\end{tabular}
\label{Tab2}
\end{table}

The residuals $x=\left\{A-\tilde{A},B-\tilde{B},C-\tilde{C}\right\}$ between the fitting parameters and the trend functions were computed, from which histograms were constructed to quantify the dispersion from each of the trends.
Employing Rice's rule [\cite{Rice}] to determinate the number of bins $M_x$ and their width $\Delta x$.
\begin{align}
\Delta x&=\dfrac{\max\left[x\right]-\min\left[x\right]}{2\left(365^{1/3}\right)} &
M_x &=\left\lceil 2\left(365^{1/3}\right)\right\rceil =15
\end{align}
The values of $\Delta x$ and $M_x$, computed from the data presented in Figure~\ref{Fig3}, are given in Table~\ref{Tab3} for each parameter.
With these values, we obtain the histograms that are also shown at the bottom of Figure~\ref{Fig3}.

\begin{table}[h]
\renewcommand\arraystretch{1.5}
\centering
\caption{Number of bins $M_x$ and width $\Delta x$ for the histograms of the residuals $x$ of each parameter $A$, $B$ and $C$ with ther corresponding general trends $\tilde{A}$, $\tilde{B}$ and $\tilde{C}$, respectively, and the distribution parameters $\mu$ and $\nu$ of the distributions $f_A$, $f_B$ and $f_C$ of the corresponding residuals $x_A$, $x_B$ and $x_C$.}
\begin{tabular}{c|c|c|c|c||c|c|c}
Normalization & \multirow{2}{*}{Trend} & Residual & No. of bins & Width & \multicolumn{3}{c}{Distribution} \\
parameter & & $x$ & $M_x$ & $\Delta x$ & $f(x)$ & $\mu$ & $\nu$ \\ \hline
$A$ & $\tilde{A}$ & $x_A=A-\tilde{A}$ & 15 & 0.0042 & $f_A(x)$ & 0.0028 & 0.0064 \\ \hline
$B$ & $\tilde{B}$ & $x_B=B-\tilde{B}$ & 15 & 0.0078 & $f_B(x)$ & 0.0048 & 0.0111 \\ \hline
$C$ & $\tilde{C}$ & $x_C=C-\tilde{C}$ & 15 & 20.7895 & $f_C(x)$ & 26.0092 & 26.0947 \\ \hline
\end{tabular}
\label{Tab3}
\end{table}

Additionally, probability distribution functions (PDFs) were found for the residuals $x=\left\{A-\tilde{A},B-\tilde{B},C-\tilde{C}\right\}$, by means of fitting procedures with the appropriated distribution shapes.
Exponential-based distributions were suggested to fulfill the PDFs requirements, due to their likeness with the corresponding histograms.
Gumbel distribution functions [\cite{Abramowitz}] had been employed :
\begin{equation}
f\left(x\right)=\dfrac{1}{\nu}\exp\left\{\dfrac{\left(x-\mu\right)}{\nu}-\exp\left[\dfrac{\left(x-\mu\right)}{\nu}\right]\right\} \ ,
\end{equation}
which expected value is $\mathbf{E}[x]=\mu-\nu\gamma$, being $\gamma\approx0.5772$ the Euler-Mascheroni constant.
Their corresponding fitting parameters are presented in Table~\ref{Tab3}, whereas their comparison and good agreement with the trends shown by the histograms are observed also at the bottom of Figure~\ref{Fig3}.
These non-symmetric PDFs present their highest probability at positive values of the residuals $x$, relative to the complete range of $x$, and a negative skewness, \emph{i.e.} the negative-resudials tail is longer.
These observations are logical, since it is more natural and usual to undergo a decrease than an increase of GHI.

\subsection{Probability maps}

The information given by the data presented in Figure~\ref{Fig4} is summarized in Figure~\ref{Fig5}, for each season.
The discrete probability maps represent the residuals from the simple parabolic master curve, obtained as it has been described in subsection~\ref{subsec:Res}.
The overlapping data presented in Figure~\ref{Fig4} is now represented as a normalized frequency, indicating the probability at which a variation of the residuals happens at a given time of the day and season.
Close to sunrise and nightfall, the most likely range of $r^{\ast}$, the rate of change of the residuals, is narrow.
In contrast, the range of $r^{\ast}$ becomes wide around noon, since the probability distribution presents long tails for both positive and negative values of $r^{ast}$.
A gradual increase of the tails starts from sunrise, reaches its maximum at noon, and gradually decreases until nightfall.
It is important to notice that for any season and any time during a day, a value $r^{\ast}\approx 0$ is the most likely to occur than any other value of $r^{\ast}$.
Nevertheless, small values $r^{\ast}\in(-0.1,0.1)$ present a signifitnact probability to happen.

The PDMs depicted in Figure~\ref{Fig6} correspond to continuous versions of the discrete maps presented in Figure~\ref{Fig5}.
These PDMs should be employed as follows:
\begin{enumerate}
\item For a given value of the short-term variable $m$ compute its dimensionless equivalent $m^{\ast}$. One must know beforehand the values of the normalization parameters $A(d)$ and $B(d)$, according to the day number.
\item According to the season, find the horizontal coordinate $m_j^{\ast}$ that is closest to the calculated value of $m^{\ast}$.
\item Along a vertical line, the vertical coordinate indicates the rate of change $r^{\ast}$ of the residuals $R^{\ast}$, recalling that $2/J=0.0313$ for the presented data.
\item Finally, recalling that herein $h=0.0364$, the color indicates the probability of $r^{\ast}$ occuring at $m^{\ast}$.
\end{enumerate}

The PDMs were transformed into scaled-PDMs by multiply it by the bandwidth $h$, only to find a normalized map, which values remain in the range $[0,1]$, and compare with the corresponding discrete probability maps.
Nevertheless, the objective of the PDMs is not only visual, since they are required as guidelines for the generation of the residuals (random numbers) for a given value of the short-term variable $m^{\ast}$ .

\section{Stochastic model description}

The daily radiant exposure $I$, in W$\cdot$h$/$m$^2$, is obtained after integration of the irradiance $E(d,m)$ with respect to the short-term variable $m$:
\begin{equation}
I(d)=\dfrac{1}{60}\int_0^{1440} E\left(d,m\right) dm \ ,
\label{Eq:Daily}
\end{equation}
where the coefficient 1$/$60 is required to perform a units conversion from minutes to hours.
\\
One may proceed to employ the irradiance model for $E(d,m)$ developed in this study, which reads:
\begin{equation}
E\left(d,m\right)=C\left(d\right)\left[\vphantom{X^X}1+E^{\ast}\left(m^{\ast}\right)\right] \ ,
\end{equation}
recalling that the normalized short-term variable $m^{\ast}$ is related to $m$ according to Eq.\eqref{Eq:NormVar}.
The normalized irradiance $E^{\ast}\left(m^{\ast}\right)$ is proposed to be decomposed into a deterministic term, the irradiance estimator $\mathcal{E}\left(m^{\ast}\right)$, and a stochastic term, the residuals $R^{\ast}$, as proposed in Eq.~\eqref{Eq:ERes}.
Also, the irradiance estimator has been introduced in Eq.~\eqref{Eq:Master}, which leads to:
\begin{equation}
E\left(d,m\right)=C\left(d\right)\left[1-\left(m^{\ast}\right)^2+R^{\ast}\left(m^{\ast}\right)\right] \ .
\label{Eq:CompMod}
\end{equation}
Additionally, according to Table~\ref{Tab3}, the parameters $A$, $B$ and $C$ can be split, as well, into the deterministic terms $\tilde{A}$, $\tilde{B}$ and $\tilde{C}$, and the stochastic terms $x_A$, $x_B$ and $x_C$, respectively, which yields:
\begin{subequations}
\begin{equation}
E\left(d,m\right)=\left[\tilde{C}\left(d\right)+x_C\right]\left[1-\left(m^{\ast}\right)^2+R^{\ast}\left(m^{\ast}\right)\right]
\end{equation}
with:
\begin{equation}
m^{\ast}=\dfrac{1}{\left[\tilde{B}\left(d\right)+x_B\right]}\left[\dfrac{m}{m_c}-\tilde{A}\left(d\right)-x_A\right] \ .
\end{equation}
\label{Eq:ESim}
\end{subequations}
This equation allows the simulation of irradiance data, by means of stochastic methods, whenever the probability distributions of the random variables $x_A$, $x_B$, $x_C$ and $r^{\ast}\left(m^{\ast}\right)$ are known.
\\
The daily radiant exposure $I(d)$ can be obtained by performing the integral in eq.~\eqref{Eq:Daily}, applied to the expresion in eq.~\eqref{Eq:CompMod}, with the change of variables $m\rightarrow m^{\ast}$ expressed in Eq.\eqref{Eq:NormVar}, and within the domain $m^{\ast}\in\left[-1,1\right]$, since $E^{\ast}=0$ is considered outside the corresponding boundaries.
Therefore, remembering that $m_c=360$, we find the following expression:
\begin{align}
I(d)
&=6\, B\left(d\right)C\left(d\right)\left\{\int_{-1}^{1} \left[1-\left(m^{\ast}\right)^2\right]\, dm^{\ast}+\int_{-1}^{1} R^{\ast}\left(m^{\ast}\right)\,dm^{\ast}\right\} \ .
\end{align}
The first integral is equal to $4/3$, whereas the second one requires a more detailed analysis, which will be given later in this section.
Additionally, according to Table~\ref{Tab3}, the parameters $B$ and $C$ can be split, as well, into the deterministic terms $\tilde{B}$ and $\tilde{C}$, and the stochastic terms $x_B$ and $x_C$, respectively, which yields:
\begin{subequations}
\begin{align}
I(d)
&=6\left[\tilde{B}\left(d\right)+x_B\right]\left[\tilde{C}\left(d\right)+x_C\right]\left[\dfrac{4}{3}+\int_{-1}^{1} R^{\ast}\left(m^{\ast}\right)\,dm^{\ast}\right] \ .
\label{Eq:Irr}
\end{align}
This equation allows the simulation of radiant exposure data, by means of stochastic methods, whenever the probability distributions of the random variables $x_B$, $x_C$ and $R^{\ast}\left(m^{\ast}\right)$ are known.
As well, the expressions that describe the parameters $B$ and $C$ must be exposed, which should display the general trend given by Eq.~\ref{Eq:Trends}, but whose coefficients should change depending on the location.

When only discrete values of $m^{\ast}_j$ are available, in order to estimate the value of the remaining integral in Eq.~\eqref{Eq:Irr}, one may implement a numerical method.
For instance, following the trapezoidal method, one finds:
\begin{align}
\int_{-1}^{1}R^{\ast}\left(m^{\ast}\right)\, dm^{\ast} &
\approx \dfrac{1}{J}\left[R^{\ast}\left(-1\right)+2\sum_{j=1}^{J-1}R^{\ast}\left(m_j^{\ast}\right)+R^{\ast}\left(1\right)\right]
\label{Eq:ResDisc}
\end{align}
\label{Eq:ISim}
\end{subequations}
according to the definition of the discrete variable $m^{\ast}_j$, given in Eq.~\eqref{Eq:Discrete}.

Finally, an average behavior is retrieved when the expected value of $I(d)$ is computed.
Assuming that all the random variables are independent, one gets:
\begin{subequations}
\begin{align}
\mathbf{E}\left[I(d)\right]
&=6\left\{\tilde{B}\left(d\right)+\mathbf{E}\left[x_B\right]\right\}\left[\tilde{C}\left(d\right)+\mathbf{E}\left[x_C\right]\right\}\left\{\dfrac{4}{3}+\int_{-1}^{1} \mathbf{E}\left[R^{\ast}\left(m^{\ast}\right)\right]\,dm^{\ast}\right\} \ .
\end{align}
The corresponding discrete approximation of the integral, using the trapezoidal method, yields:
\begin{align}
\int_{-1}^{1}\mathbf{E}\left[R^{\ast}\left(m^{\ast}\right)\right]\, dm^{\ast} &
\approx \dfrac{1}{J}\left\{\mathbf{E}\left[R^{\ast}\left(-1\right)\right]+2\sum_{j=1}^{J-1}\mathbf{E}\left[R^{\ast}\left(m_j^{\ast}\right)\right]+\mathbf{E}\left[R^{\ast}\left(1\right)\right]\right\}
\end{align}
\label{Eq:ExpI}
\end{subequations}

The expected value of the residual $R^{\ast}$ at a given value of the discrete short-term variable $m^{\ast}_j$ should be obtained directly from the integral:
\begin{equation}
\mathbf{E}\left[R^{\ast}\left(m_j^{\ast}\right)\right]=\left\{\int_{-\infty}^{\infty} R^{\ast}g\left[R^{\ast}\right]dR^{\ast}\right\}_{m^{\ast}_j} \ ,
\end{equation}
where $g\left[R^{\ast}\right]$ is the probability density function of $R^{\ast}$.
The generation of random values of $R^{\ast}$, recalling that $R^{\ast}$ depends on the independent random variable $r^{\ast}$, requires the following procedure.
From eq.~\eqref{Eq:resdif}, and applying once more the trapezoidal integration method, a discrete version in a recursive form reads:
\begin{equation}
R^{\ast}\left(m^{\ast}_j\right)=R^{\ast}\left(m^{\ast}_{j-1}\right)+\left(\dfrac{2}{J}\right) r^{\ast}\left(m^{\ast}_j\right) \ .
\end{equation}
Therefore, a random value of $R^{\ast}\left(m^{\ast}_j\right)$ is obtained from its forerunner $R^{\ast}\left(m^{\ast}_{j-1}\right)$ and a randomly generated value of $r^{\ast}\left(m^{\ast}_j\right)$.
This process is performed ensuring that the generated value of $R^{\ast}$ does not trespass the boundaries fixed by the criterion $f_h[r^{\ast}]>0.001$ for the corresponding $m^{\ast}_j$, which indicate the maximum and minimum allowable values of $R^{\ast}\left(m^{\ast}_j\right)$.

\begin{figure}[!h]
\centering
\includegraphics[width=\textwidth]{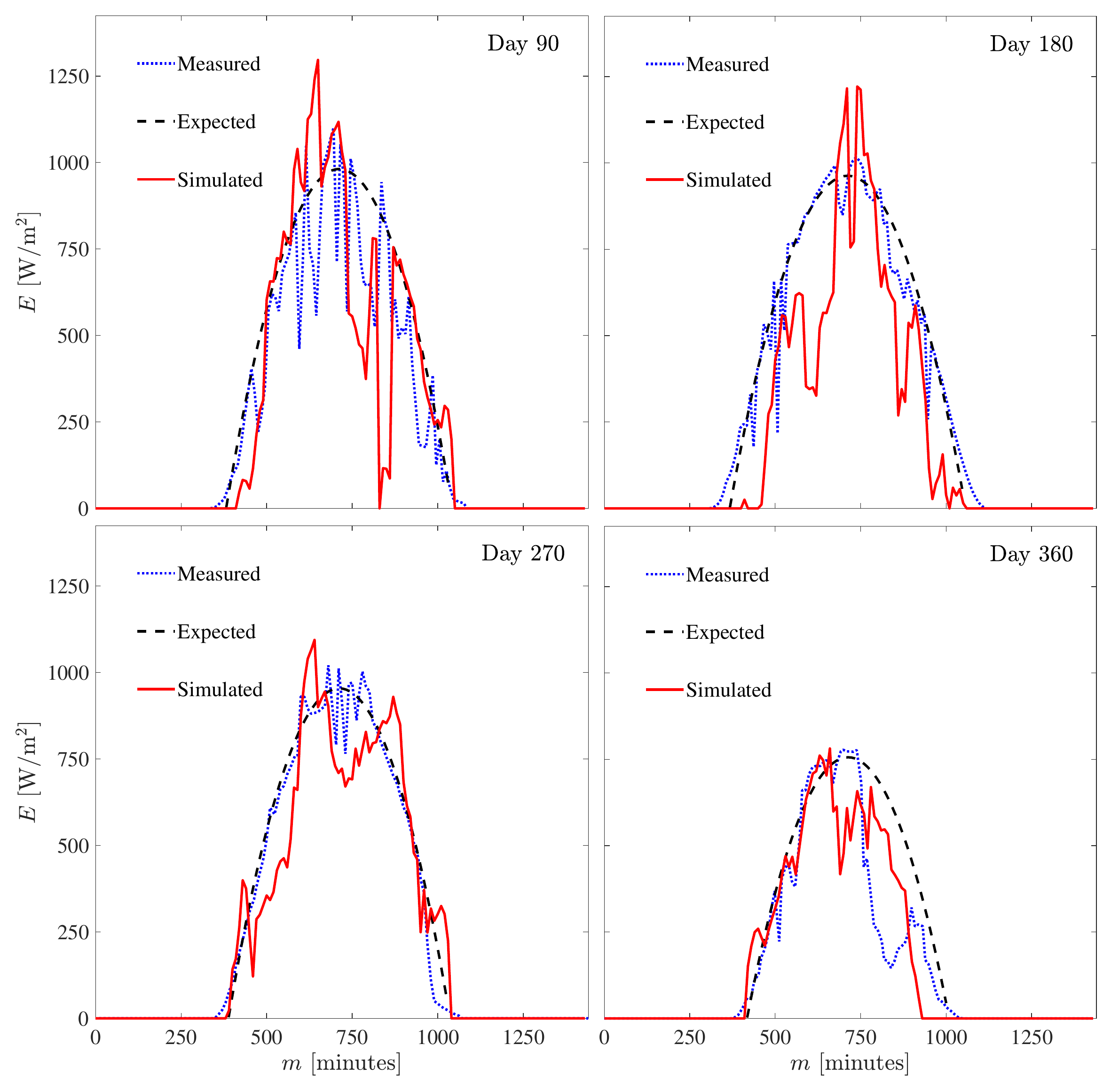} 
\caption{Irradiance profiles, measured (dotted and blue), expected (dashed and black) and simulated (solid and red), for four different dates of a year, as described in Table~\ref{Tab1}..}
\label{Fig7}
\end{figure}

\begin{figure}[!h]
\centering
\includegraphics[width=\textwidth]{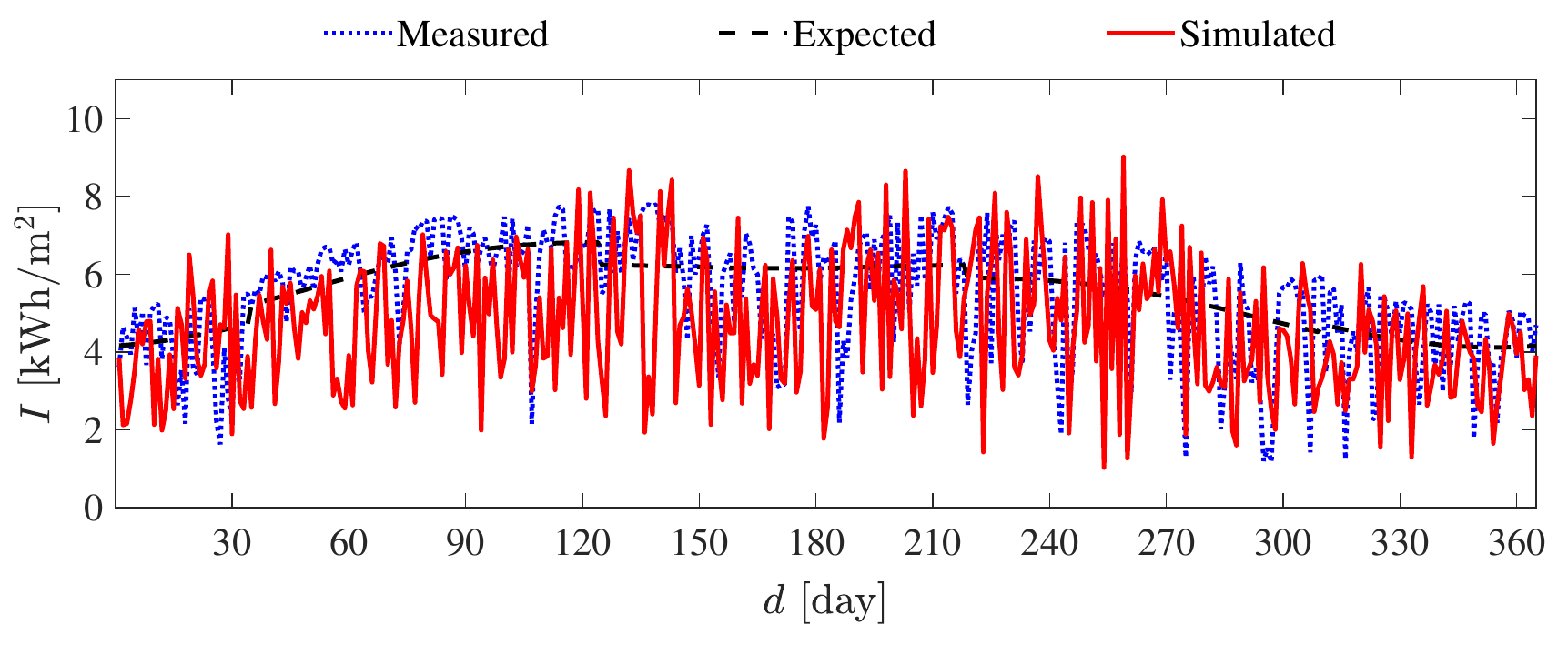} 
\caption{Daily radiant exposure, measured (dotted and blue), expected (dashed and black) and simulated (solid and red), along the year described in Table~\ref{Tab1}.}
\label{Fig8}
\end{figure}

We employ Eqs.~\eqref{Eq:ESim} to simulate the irradiance $E$ throughout the day, whereas Eq.~\eqref{Eq:ISim} allows us to simulate the radiant exposure $I$ for an entire day.
The random values of $x_A$, $x_B$, $x_C$ and $r^{\ast}(m^{\ast})$ had been generated by means of the inverse transform sampling method.
Examples of the irradiance $E$ (measured data, expectation and simulation) as a function of the short-term variable $m$, for four dates, are presented in Fig.~\ref{Fig7}, whereas the daily radiant exposure $I$ (measured data, expectation and simulation) as a function of the long-term variable $d$, for a year, is presented in Fig.~\ref{Fig8}.

\section{Validation tests}
\label{Sec:Sims}

Once our empiric-stochastic model was completed, some validation tests were made to check its correct performance.
The first validation test was performed to check the short-term reliability, simulations of solar irradiance over a day, whereas the second validation test was done to verify the long-term behavior, simulations of the radiant exposure over a year.

The firts test consisted in a comparisons between the measured, the expected and the simulated irradiance profiles $E(m)$.
Examples are presented in Figure~\ref{Fig7}, for four different dates of the year (days 90, 180, 270 and 360).
The measured data corresponds to the data depicted in Figure~\ref{Fig1}, the expected behaviors were computed using eqs.~\eqref{Eq:Fit} and the corresponding normalization parameters, and the simulated profiles were constructed with eq.~\eqref{Eq:ESim} and the random generation of the corresponding parameters.
All the procedures were described in the previous section.
For the cases presented for all seasons, the expected behavior describes the trends of measurements and simulations, whereas a qualitative resemblance is observed between the measured and simulated data, which is an important objective of our model.

The second test consisted in a comparison between the measured, the expected and the simulated radiant exposures $I(d)$.
An example is presented in Figure~\ref{Fig8}, for a year.
The measured data corresponds to the integral, defined in eq.~\eqref{Eq:Daily} and applied to each irradiance curve $E(m)$ in Figure~\ref{Fig1}, corresponding to a different day $d$.
The expected behavior was obtained from eqs.~\eqref{Eq:ExpI} and the expected values of the corresponding  probability distributions.
The simulated data was constructed with eq.~\eqref{Eq:ISim} and the random generation of the residuals.
All the procedures were described in the previous section.
A qualitative likeness between experimental and simulated curves is observed, whereas the expected curve clearly indicates the common trend.

\begin{figure}[!h]
\centering
\includegraphics[width=\textwidth]{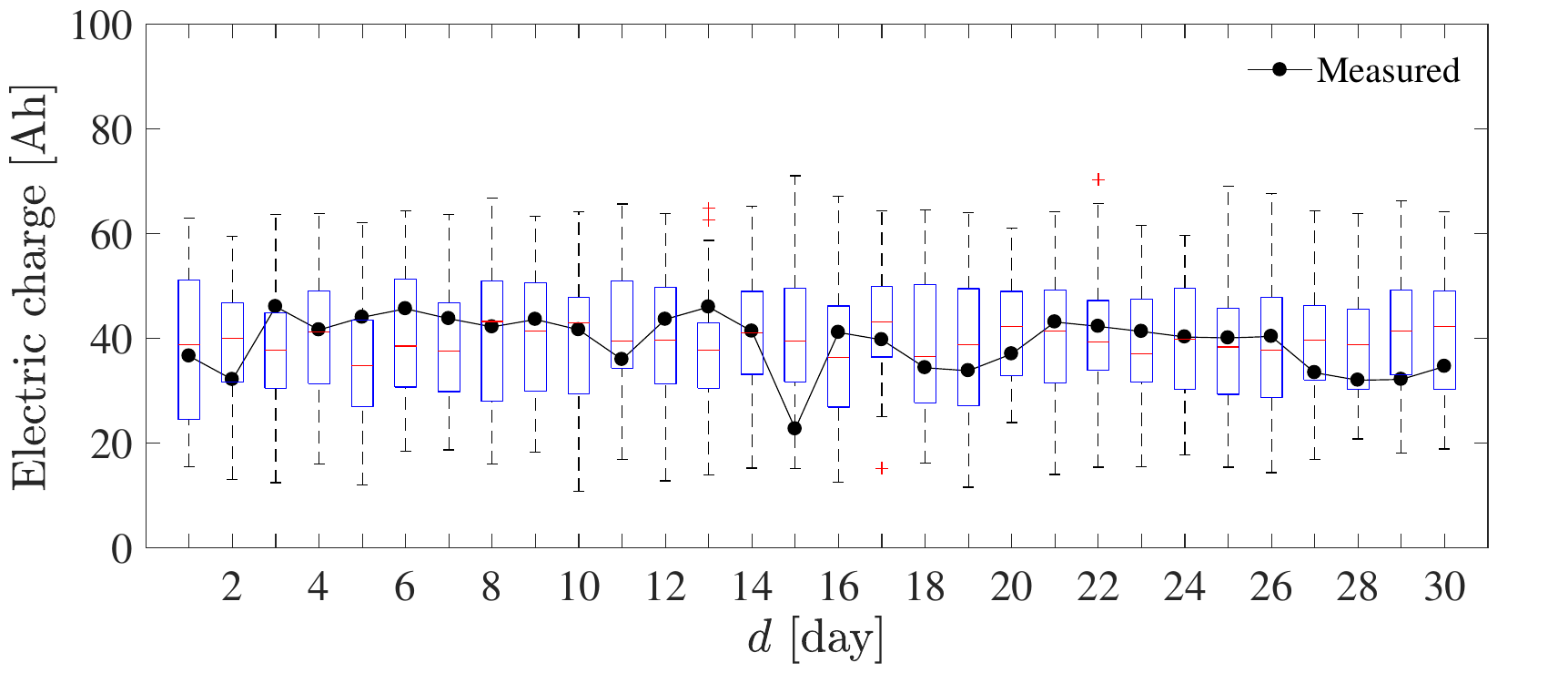} 
\caption{Comparison of the measurements obtained from a photovoltaic system (black circles and line) with the statistics obtained from the simulation of a year, with one hundred samples. The (blue) boxes, the (red) dashes, the (red) crosses and the (dashed) whiskers indicate the quantiles, the median, the outliers and the full range of the data, respectively.}
\label{Fig9}
\end{figure}

Despite the good qualitative agreement and the good description of the average behaviors, a further validation test was performed.
The comparison between the electric charge, generated with a photovoltaic system (PV) system throughout a month, and the statistics of simulated electric charge was done.
On the one hand, the PV system is installed at the UQRoo-Chetumal campus, and consists of a series-connected array of eight solar panels, which characteristics are enlisted in Table~\ref{Tab4}.
On the other hand, the statistics were obtained from a hundred simulations, which were performed for each day and for the aforementioned period of time, a month.
Each simulation was employed as the input of a single-diode solar-cell model [\cite{Saloux2011,Ma2014,Chaibi2018}], to generate electric current data as the output.
Subsequently, the electric current along a day was integrated over time to obtain the electric charge, for each of the hundred simulations of each day.
Finally, some statistics were computed, \emph{i.e.} median, quantiles and full range, for each day $d$.
The comparison between the measured data and the simulated statistics can be observed in figure~\ref{Fig9}, where the quantiles and the full range of the simulated data are depicted as a box and whiskers, respectively.
The measured data remains within the limits of the quantile boxes for most of the days, which indicates that the individual simulations should be qualitatively equivalent to the measured data and will provide a good prediction of the electric charge that can be generated along a month.

\begin{table}[h]
\renewcommand\arraystretch{1.0}
\centering
\caption{Nominal specifications of the Solartec\texttrademark\, S60PC-250 solar panel  [\cite{S60PC}].\\
\small STC - Standard Test Conditions; PTC - Photovoltaics for Utility Scale Applications Test Conditions; NOCT - Nominal Operating Cell Temperature}
\begin{tabular}{c|c}
Parameter & {Value} \\ \hline
STC Power & 250 W  \\ \hline
PTC Power & 226.47 W  \\ \hline
NOCT & 45$^{\circ}$ \\ \hline
SCT Power per unit area & 153.7 W$/$m$^2$ \\ \hline
Peak efficiency & 15.39 \% \\ \hline
Number of cells & 60 \\ \hline
Maximum power current (Imp) & 8.17 A \\ \hline
Maximum power voltage (Vmp) & 30.60 V  \\ \hline
Short circuit current (Isc) & 8.71 A \\ \hline
Open circuit voltage (Voc) & 36.3 V \\
\end{tabular}
\label{Tab4}
\end{table}

\section{Conclusions}
\label{Sec:Conclusions}

A hybrid approach to generate global horizontal irradiantion was introduced.
For a given location, solar irradiance data can be employed to compute normalization parameters, average trends and probablity distributions, in order to gather the essential ingredients of the methodology.
Once these information is known, we proceed to generate random numbers, according to the probability distributions, that are added to the average trends and subjected to a de-normalization procedure, to retrieve solar irradiance data, simulated for a single daytime or a complete year.

The proposed method was tested and validated with a direct and an undirect approach, by comparison with solar irradiance data, measured by a metheorological station, and electric charge, generated by a photovoltaic solar system.
A good qualitative agreement between experiments and simulations was found.

For a given location, some experimental data is required to generate the parameters that the methodology employs.
Nevertheless, the proposed methodology allows the implementation of an alternative inverse engineering sequence.
An educated guess of the normalization parameters, average trends and probability distributions can be done, and afterwards, compare the results against a minimum amunt of experimental data.

The presented methodology is presented as an alternative and simple procedure to provide useful data , which together with an energy load profile, can be employed as the input of a succeeding analysis, for instance a simulation software for the control and surveillance of an intelligent energy network.








%
\end{document}